\documentclass[twocolumn,letterpaper,aps,prl,superscriptaddress,showpacs,floatfix]{revtex4-1}

\usepackage{amsmath}
\usepackage{graphicx}	

\usepackage{xspace}	

\newcommand{\sqsn}{\mbox{$\sqrt{s_{_{NN}}}$}\xspace}
\newcommand{\lt}{<}
\newcommand{\gt}{>}
\begin{document}

%
\title{ Observation of the critical end point in the phase diagram for \\ hot and dense nuclear matter}

%

\author{Roy A. Lacey}
\address{Depts. of Chemistry \& Physics, Stony Brook University, NY 11794}

\date{\today}

\begin{abstract}
Excitation functions for the Gaussian emission source radii difference ($R^2_{\text{out}} - R^2_{\text{side}}$) 
obtained from  two-pion interferometry measurements in 
Au+Au ($\sqrt{s_{NN}}= 7.7 - 200$ GeV)  and Pb+Pb ($\sqrt{s_{NN}}= 2.76$ TeV) collisions, 
are studied for a broad range of collision centralities. The observed non-monotonic excitation functions 
validate the finite-size scaling patterns expected for the deconfinement phase transition and 
the critical end point (CEP), in the temperature  vs. baryon chemical potential ($T,\mu_B$) plane
of the nuclear matter phase diagram. 
A Finite-Size Scaling (FSS) analysis of these data indicate a second order phase transition with 
the estimates $T^{\text{cep}} \sim 165$~MeV and $\mu_B^{\text{cep}} \sim 95$~MeV 
 for the location of the critical end point. The critical exponents ($\nu \sim 0.66$ and $\gamma \sim 1.2$)
extracted via the same FSS analysis, places the CEP in the 3D Ising model universality class.

\end{abstract}

	



\maketitle
 

%
One of the most fundamental phase transitions is that between the hadron gas and the 
Quark Gluon Plasma (QGP). This Deconfinement Phase Transition (DPT) is usually
depicted in the plane of temperature vs. baryon chemical potential ($T, \mu_B$)
in the conjectured phase diagram for Quantum Chromodynamics (QCD)
 \cite{Itoh:1970,Shuryak:1983zb,Asakawa:1989bq,Stephanov:1998dy}.
The detailed character of this QCD phase diagram is not known and current theoretical knowledge 
is restricted primarily to the $\mu_B=0$ axis.  

Lattice QCD calculations indicate a crossover quark-hadron 
transition at small $\mu_{B}$ or high collision energies (\sqsn)~\cite{Aoki:2006we,Bhattacharya:2014ara}. 
Similar calculations for much larger $\mu_B$ values have been hindered by the well known {\it sign problem}~\cite{deForcrand:2010ys}.
However, several model approaches~\cite{Berges:1998rc,Hatta:2002sj,Stephanov:2004wx,Asakawa:2005hw,Ejiri:2008xt},
as well as mathematical extensions of lattice techniques~\cite{Fodor:2004nz,Li:2011ee,Nagata:2014fra,deForcrand:2014tha}, 
indicate that the transition at larger values of $\mu_{B}$ (lower beam energies~\cite{chemfo}) is strongly
first order, suggesting the existence of  a critical end point (CEP). Pinpointing the location of the phase 
boundaries and the CEP is central  to ongoing efforts to map the QCD phase diagram 
and to understand the properties of strongly interacting matter under extreme conditions.

The matter produced in ultrarelativistic heavy ion collisions can serve as an 
important  probe for the phase boundaries and the 
CEP \cite{Itoh:1970,Shuryak:1983zb,Asakawa:1989bq,Stephanov:1998dy}.   
Indeed, a current experimental strategy at the Relativistic Heavy Ion Collider (RHIC) is centered 
on beam energy scans which enable a search  for non-monotonic excitation functions over a broad  
domain of  the ($T, \mu_B$)-plane. 
The rationale is that the expansion dynamics of the matter produced  in these beam energy scans, is strongly influenced 
by the path of the associated reaction trajectories in the  ($T, \mu_{B}$)-plane.  Trajectories which 
are close to the CEP or cross the coexistence curve for the first order phase transition, are expected 
to be influenced by anomalies in the dynamic properties of the medium. Such anomalies can drive abrupt 
changes in the transport coefficients and relaxation rates to give a non-monotonic dependence 
of  the excitation function for the specific viscosity $\frac{\eta}{s}$ {\em i.e.} the ratio of the shear viscosity $\eta$ 
to entropy density $s$~\cite{Lacey:2006bc,Csernai:2006zz,Lacey:2007na}.

An emitting system produced in the vicinity of the CEP would also be subject to the influence of 
a divergence in the compressibility of the medium, resulting in  a precipitous  drop in the sound speed
and a collateral increase in the emission duration. Such effects could also give rise to non-monotonic dependencies 
in the excitation functions for the expansion speed~\cite{Hung:1994eq,Rischke:1996em}, as well as for the 
difference between the Gaussian emission source radii ($R^2_{\text{out}} - R^2_{\text{side}}$) extracted from  two-pion 
interferometry measurements \cite{Pratt:1984su,Hung:1994eq,Chapman:1994yv,Wiedemann:1995au,Rischke:1996em}.
The latter is linked to the emission duration.

In recent work~\cite{Lacey:2013qua,Lacey:2014rxa},  a striking pattern of viscous damping, compatible with the 
expected minimum in the excitation function for $\frac{\eta}{s}$~\cite{Csernai:2006zz,Lacey:2007na}
was reported for Au+Au ($\sqrt{s_{NN}}= 7.7 - 200$ GeV)  and Pb+Pb ($\sqrt{s_{NN}}= 2.76$ TeV) collisions.
An excitation function for ($R^2_{\text{out}} - R^2_{\text{side}}$) extracted  for central collisions from the same 
data sets, also indicated a striking non-monotonic pattern  attributed to decay
trajectories close to the CEP~\cite{Lacey:2014rxa,Adare:2014qvs}. Nonetheless, it remains 
a crucial open question as to whether these non-monotonic patterns are indeed linked to the 
deconfinement phase transition and the CEP?
%
%
\begin{figure*}[t]
\includegraphics[width=1.0\linewidth]{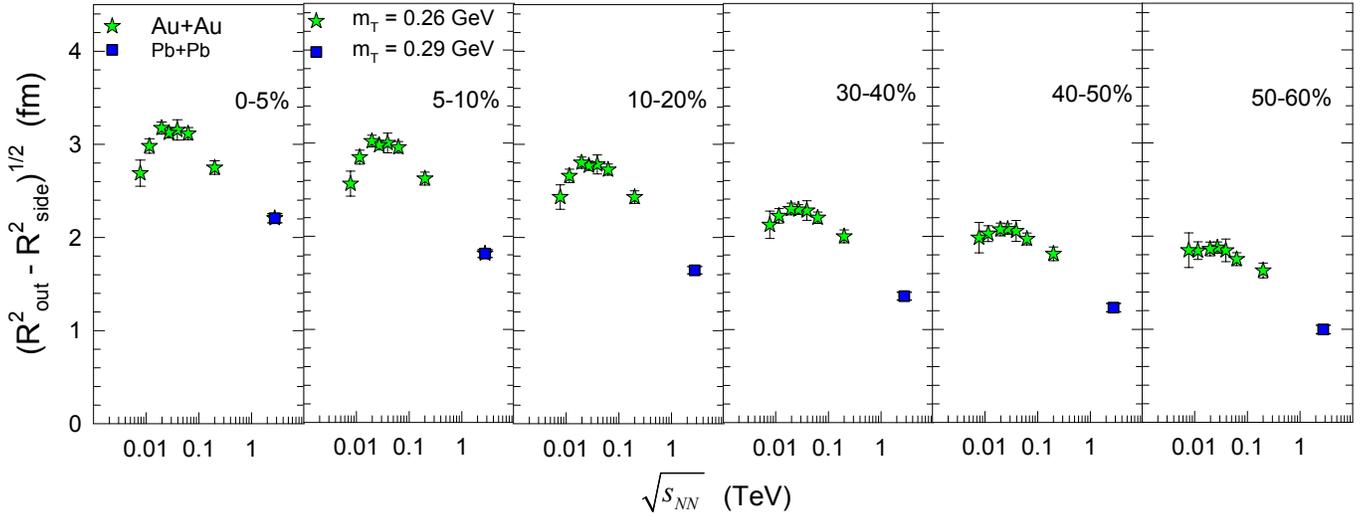}
\vspace{-0.4cm}
\caption{(Color online) $\sqrt{(R^2_{\text{out}} - R^2_{\text{side}})}$ vs. $\sqrt{s_{NN}}$  
for 0-5\%,  5-10\%, 10-20\%, 30-40\%, 40-50\% and 50-60\% Au+Au  and Pb+Pb collisions
for $m_T = 0.26$~GeV and 0.29~GeV respectively. The data are taken from 
Refs. \cite{Adamczyk:2014mxp,Aamodt:2011mr,Kisiel:2011jt}
}
\label{Fig1}
\vspace{-4pt}
\end{figure*}

In the limit of an infinite volume, the deconfinement phase transition  is characterized by singularities
which reflect the divergences in the derivatives of the thermodynamic potential, eg., the specific heat 
and various susceptibilities ($\chi$). Discontinuities in the first and second derivatives signal 
the first order and  second order phase transitions respectively. 
These singularities are smeared into finite peaks with modified positions and widths, 
for more restricted volumes~\cite{Ladrem:2004dw,Palhares:2009tf}.

The correlation length $\xi$ diverges near the transition temperature ($T^{\text{cep}}$) 
as $\xi \propto \left| \tau \right|^{-\nu}$ for an infinite volume; $\tau = T - T^{\text{cep}}$. However, 
for a system of size $L^d$ ($d$ is the dimension) this second order phase transition is expected to 
show a pseudocritical point for correlation length $\xi \approx L$. This leads to 
a characteristic power law volume (V) dependence of the magnitude ($\chi ^{\text{max}}_{T}$), 
width ($\delta T$) and peak position ($\tau_T$) of the susceptibility \cite{Ladrem:2004dw};
\begin{eqnarray}
\chi ^{\text{max}}_T(V)  \sim L^{\gamma /\nu},          
\label{eq:1} \\
\delta T (V)  \sim  L^{- \frac{1}{\nu}},                                                                               
\label{eq:2} \\
\tau_T(V) \sim  T^{\text{cep}}(V) - T^{\text{cep}}(\infty)   \sim  L^{- \frac{1}{\nu}},  
\label{eq:3}
\end{eqnarray}
where $\nu$ and $\gamma$ are critical exponents which characterize the divergence of 
$\xi$ and $\chi_T$ respectively. The reduction of the magnitude of  $\chi ^{\text{max}}_T(V)$
($\chi ^{\text{max}}_{\mu_B}(V)$ ), broadening of the transition region $\delta T (V)$ ($\delta \mu_B (V)$)
and the shift of  $T^{\text{cep}}$ ($\mu_B^{\text{cep}}$) increases as the volume decreases.
A similar set of volume or finite-size dependencies is expected for the first order 
phase transition, but with unit magnitudes for the critical exponents~\cite{Ladrem:2004dw}. Thus, a profitable route 
for locating the CEP is to search for, and utilize the characteristic finite-size scaling patterns associated 
with the deconfinement phase transition~\cite{Ladrem:2004dw,Palhares:2009tf}.

In this Letter, we use the Gaussian radii ($R_{\text{out}}$ and  $R_{\text{side}}$) extracted from 
two-pion interferometry measurements, to first construct non-monotonic excitation functions 
for ($R^2_{\text{out}} - R^2_{\text{side}}$)  as a function of collision centrality. 
We then use them to perform validation tests for the characteristic finite-size scaling patterns commonly associated 
with the deconfinement phase transition and the CEP. We find clear evidence for these scaling 
properties and use a Finite-Size Scaling (FSS) analysis to extract  initial estimates  for the 
($T, \mu_B$) location of the CEP and the critical exponents associated with it.
 
The data employed in the present analysis are taken from interferometry measurements by the 
STAR collaborations for Au+Au collisions spanning the range $\sqrt{s_{NN}}= 7.7 - 200$ GeV~\cite{Adamczyk:2014mxp},
and by the ALICE collaboration for Pb+Pb collisions at $\sqrt{s_{NN}}$ = 2.76 TeV~ \cite{Aamodt:2011mr,Kisiel:2011jt}.
The STAR measurements have been reported to be in very good agreement with similar PHENIX measurements 
obtained at  $\sqrt{s_{NN}}= 39, 62.4$ and $200$~GeV~\cite{Lacey:2014rxa,Adare:2014qvs}. 
The systematic uncertainties for these measurements are also reported to be 
relatively small~\cite{Adamczyk:2014mxp,Aamodt:2011mr,Kisiel:2011jt,Adare:2014qvs}. 

The geometric quantities employed in our Finite-Size Scaling analysis 
were obtained from a Monte Carlo Glauber (MC-Glauber) calculation \cite{glauber,Lacey:2010hw,Adare:2013nff},
performed for several collision centralities at each beam energy. In each of these calculations, a subset of the nucleons 
become participants ($N_{\text{part}}$) in each collision by undergoing an initial inelastic N+N interaction.  
The transverse distribution of these participants in the X-Y  plane has RMS widths $\sigma_x$ and $\sigma_y$ 
along its principal axes. We define and compute $\bar{R}$, the characteristic initial transverse size, as 
${1}/{\bar{R}}~=~\sqrt{\left({1}/{\sigma_x^2}+{1}/{\sigma_y^2}\right)}$ \cite{Bhalerao:2005mm}.
The systematic uncertainties for $\bar{R}$, obtained via variation of the model parameters, are 
less than 10\% \cite{Lacey:2010hw,Adare:2013nff}. 

\begin{figure}[t]
\includegraphics[width=0.9\linewidth]{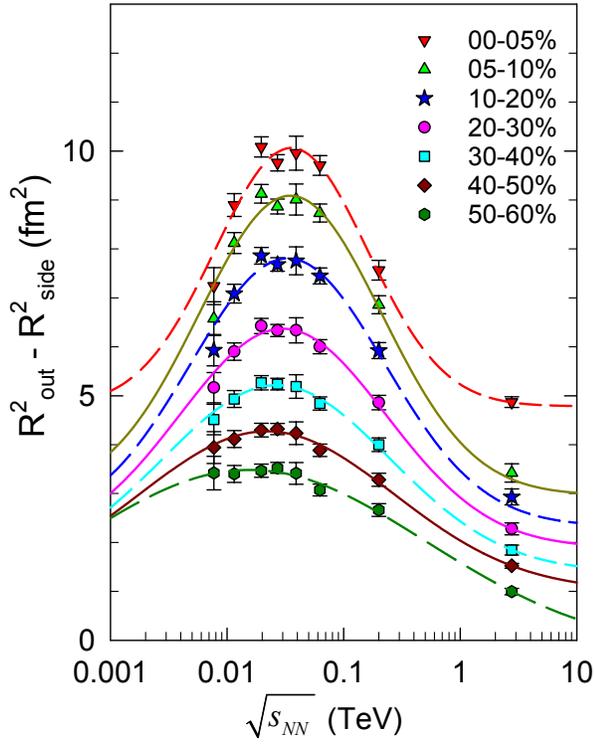}
\caption{(Color online) Comparison of ${(R^2_{\text{out}} - R^2_{\text{side}})}$ vs. $\sqrt{s_{NN}}$
for several centrality selections as indicated. The data, which are the same as those shown in Fig.~\ref{Fig1},
are taken from Refs. \cite{Adamczyk:2014mxp,Aamodt:2011mr}.
The solid and dashed curves represent Gaussian fits to the combined data sets for each centrality.
}
\label{Fig2}
\end{figure}

Figure~\ref{Fig1} shows a representative set of  excitation functions for $\sqrt{(R^2_{\text{out}} - R^2_{\text{side}})}$,
obtained for the broad selection of centrality cuts indicated.  These excitation functions, which are linked to the compressibility
of the medium, all show the non-monotonic dependence previously conjectured to reflect reaction trajectories 
close to the critical end point~\cite{Lacey:2014rxa,Adare:2014qvs}. They also exhibit several 
characteristic trends: (i) the magnitude of the peaks decrease with increasing centrality or decreasing transverse size, 
(ii) the positions of the peaks shift to lower values of $\sqrt{s_{NN}}$ with an increase in centrality and 
(iii) the width of the distributions grow with centrality. These trends are made more transparent in 
Fig.~\ref{Fig2} where a direct comparison of the excitation functions 
for ${(R^2_{\text{out}} - R^2_{\text{side}})}$ is shown. We attribute these qualitative 
patterns to  the finite-size scaling effects expected  for the deconfinement phase 
transition (cf. Eqs.~\ref{eq:1} - \ref{eq:3})  and employ the excitation functions in a 
more quantitative Finite-Size Scaling (FSS) analysis  as discussed below.

%
%

Validation tests for  finite-size scaling were carried out for the full set of excitation functions
as follows. First, we exploit the phenomenology of  thermal 
models \cite{Cleymans:2006qe,Andronic:2009qf,Becattini:2012sq,Tawfik:2014eba} for the freeze-out
region and associate ($T, \mu_B$) combinations with $\sqrt{s_{NN}}$. Second, we associate  
${(R^2_{\text{out}} - R^2_{\text{side}})}$  with a susceptibility, given its 
connection to the compressibility. Subsequently, a Guassian fit was used to extract 
the peak positions, heights and widths of the excitation functions, for different system sizes 
characterized by the centrality selections indicated in Fig.~\ref{Fig2}. The solid and dashed curves 
shown in  the figure gives an indication of the quality of these fits.  

The extracted fit parameters were tested for the characteristic finite-size scaling patterns associated with 
the deconfinement phase transition via Eqs.~\ref{eq:1} and \ref{eq:3} with $L = \bar{R}$;
\begin{eqnarray}
{(R^2_{\text{out}} - R^2_{\text{side}})}^{\text{max}} \propto \bar{R}^{\gamma /\nu},
\label{eq:4}\\
\sqrt{s_{NN}}(V)  =  \sqrt{s_{NN}}(\infty)  - k\times \bar{R}^{- \frac{1}{\nu}},
\label{eq:5}
\end{eqnarray}
with the aim of obtaining initial estimates for the critical exponents $\nu$ and $\gamma$
and the $\sqrt{s_{NN}}$ value where the deconfinement phase transition first occurs;
$k$ is constant.

\begin{figure}[t]
\includegraphics[width=1.0\linewidth]{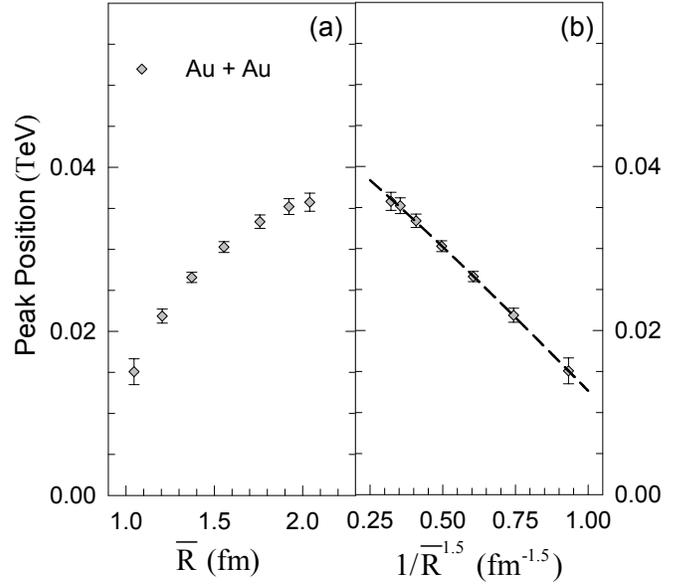}
\caption{(Color online) (a) Peak position vs. $\bar{R}$. (b) Peak position vs. $1/{\bar{R}}^{1.5}$.
The peak positions are obtained from the Gaussian fits shown in Fig.~\ref{Fig2}.
The dashed curve in (b) shows the fit to the data in (a).
}
\label{Fig3}
\end{figure}

Figure~\ref{Fig3} illustrates the finite-size scaling test made for the extracted peak positions ($\sqrt{s_{NN}}(V)$). 
Panel (a) shows the peak positions vs. $\bar{R}$ while panel (b) shows the same peak positions vs. $1/\bar{R}^{1.5}$. 
The dashed curve in (b), which represents a fit to the data in (a) with Eq.~\ref{eq:5}, confirms the expected inverse power law dependence 
of these peaks on $\bar{R}$. The fit gives the values $\sqrt{s_{NN}}(\infty)  \sim  47.5$~ GeV and $\nu \sim 0.66$.
Note that this value of $\sqrt{s_{NN}}(\infty)$ is compatible with the striking pattern observed in the 
excitation function for viscous damping~\cite{Lacey:2013qua,Lacey:2014rxa}. This pattern is akin to that 
expected for $\frac{\eta}{s}(T,\mu_B)$ close to the CEP~\cite{Csernai:2006zz,Lacey:2007na}.

%
%

\begin{figure}[t]
\includegraphics[width=1.0\linewidth]{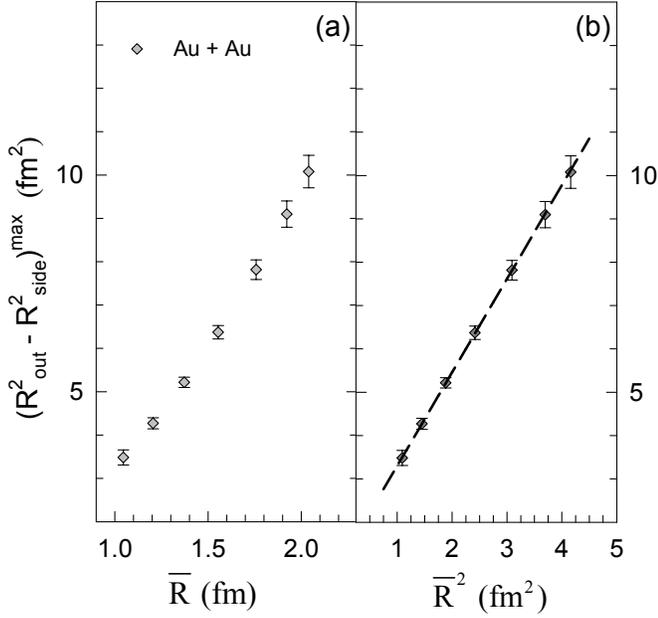}
\caption{(Color online) (a) $(R^2_{\text{out}} - R^2_{\text{side}})^{\text{max}}$ vs. $\bar{R}$.
(b)  $(R^2_{\text{out}} - R^2_{\text{side}})^{\text{max}}$ vs. $\bar{R}^2$.
The $(R^2_{\text{out}} - R^2_{\text{side}})^{\text{max}}$ values are obtained from the 
Gaussian fits shown in Fig.~\ref{Fig2}. The dashed curve in (b) shows the fit to the data in (a).
}
\label{Fig4}
\end{figure}

Figure~\ref{Fig4} illustrates the results of the finite-size scaling test for ${(R^2_{\text{out}} - R^2_{\text{side}})}^{\text{max}}$. 
Panel (a) shows ${(R^2_{\text{out}} - R^2_{\text{side}})}^{\text{max}}$  vs. $\bar{R}$ while 
panel (b) shows the same data plotted vs. $\bar{R}^{2}$. The dashed curve in (b), which represents a fit to the data in (a) with Eq.~\ref{eq:4},
confirms the expected power law dependence of ${(R^2_{\text{out}} - R^2_{\text{side}})}^{\text{max}}$ on $\bar{R}$.
Note that the trend of this dependence is opposite to the inverse power dependence shown in Fig.~\ref{Fig3}. The fit
leads to the estimate $\gamma \sim 1.2$ .

The magnitudes of the extracted values for the critical exponents $\nu \sim 0.66$ and $\gamma \sim 1.2$, are different from
the unit values expected for a first order phase transition~\cite{Ladrem:2004dw}. However, they are compatible with the 
critical exponents for the second order deconfinement phase transition for the 3D Ising model universality class~\cite{3DIsing1,3DIsing2}. 
Consequently, we assign the location of the CEP to the extracted value $\sqrt{s_{NN}}(\infty)  \sim  47.5$~GeV 
and use the parametrization for chemical freeze-out in Ref.~\cite{Cleymans:2006qe} to obtain the 
estimates $\mu_B^{\text{cep}} \sim 95$~MeV and $T^{\text{cep}} \sim 165$~MeV for its 
location in the ($T, \mu_B$)-plane.
%
%

A crucial crosscheck for the location of the CEP and its associated critical exponents, is the requirement that 
finite-size scaling for different transverse sizes, should lead to data collapse onto a single curve 
for robust values of $T^{\text{cep}}$,  $\mu_B^{\text{cep}}$ and the critical exponents $\nu$ and $\gamma$;
\begin{eqnarray}
\bar{R}^{-\gamma/\nu}\times (R^2_{\text{out}} - R^2_{\text{side}})  \text{  vs.  }  \bar{R}^{1/\nu} \times t_T,  \nonumber \\
\bar{R}^{-\gamma/\nu}\times (R^2_{\text{out}} - R^2_{\text{side}})  \text{  vs.  }  \bar{R}^{1/\nu} \times t_{\mu_B}, \nonumber
\end{eqnarray}
where $t_T = (T - T^{\text{cep}})/T^{\text{cep}}$ and  $t_{\mu_B} = (\mu_B - \mu_B^{\text{cep}})/\mu_B^{\text{cep}}$
are the reduced temperature and baryon chemical potential respectively.

\begin{figure}[t]
\includegraphics[width=1.0\linewidth]{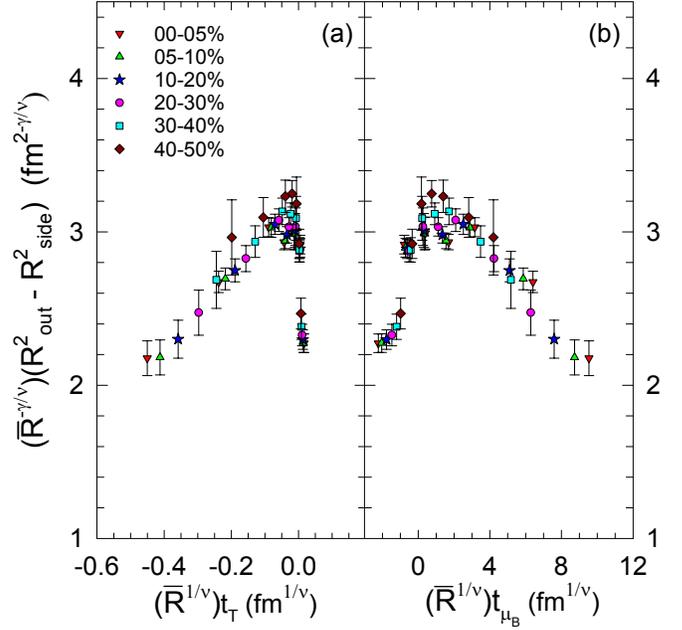}
\caption{(Color online) (a) $\bar{R}^{-\gamma/\nu} \times (R^2_{\text{out}} - R^2_{\text{side}})$ vs. $\bar{R}^{1/\nu} \times t_T$.
(b)  $\bar{R}^{-\gamma/\nu}\times (R^2_{\text{out}} - R^2_{\text{side}})$ vs. $ \bar{R}^{1/\nu} \times t_{\mu_B}$.
The $(R^2_{\text{out}} - R^2_{\text{side}})$ values are the same as those in Fig.~\ref{Fig2}. 
The parametrization for chemical freeze-out \cite{Cleymans:2006qe} is used in conjunction with  
$\mu_B^{\text{cep}}$ and  $T^{\text{cep}}$ to determine $t_T$ and  $t_{\mu_B}$.
}
\label{Fig5}
\end{figure}

The validation of this crosscheck is illustrated in Fig.~\ref{Fig5} where data collapse onto a single curve is indicated 
for the RHIC excitation functions shown in Fig.~\ref{Fig2}.  
The parametrization for chemical freeze-out \cite{Cleymans:2006qe} is used in conjunction with  $\mu_B^{\text{cep}}$ 
and  $T^{\text{cep}}$ to determine the required $t_T$ and  $t_{\mu_B}$ values from  the $\sqrt{s_{NN}}$ values 
plotted in Fig.~\ref{Fig2}. Figs.~\ref{Fig5}(a) and (b) also validate the expected  trends for reaction trajectories 
in the ($T, \mu_B$) domain  which encompass the CEP. 
That is, the scaled values of $(R^2_{\text{out}} - R^2_{\text{side}})$ peaks at $t_T \sim 0$ and $t_{\mu_B} \sim 0$, 
and show the collateral fall-off  for $t_{T, \mu_B} \lt 0$ and $t_{T, \mu_B} \gt 0$.

%
%
%

%
In summary, we have investigated  the centrality dependent excitation functions for the Gaussian emission 
source radii difference ($R^2_{\text{out}} - R^2_{\text{side}}$), obtained from  two-pion interferometry 
measurements in Au+Au ($\sqrt{s_{NN}}= 7.7 - 200$ GeV)  and Pb+Pb ($\sqrt{s_{NN}}= 2.76$ TeV) 
collisions, to search for the CEP in the nuclear matter phase diagram. 
The observed centrality dependent  non-monotonic excitation functions, validate the characteristic 
finite-size scaling patterns expected for the deconfinement phase transition and the critical end point.
An initial  Finite-Size Scaling analysis of these data suggest a second order phase transition with 
$T^{\text{cep}} \sim 165$~MeV and $\mu_B^{\text{cep}} \sim 95$~MeV 
 for the location of the critical end point. The critical exponents ($\nu \sim 0.66$ and $\gamma \sim 1.2$)
extracted in the same FSS analysis, places the CEP in the 3D Ising model universality class.  
Further detailed studies at RHIC are crucial to make a more precise determination of the location of the 
CEP and the associated critical exponents, as well as to confirm these observations  for 
other collision systems.

\section*{Acknowledgments}
The author thanks J. Jia, E. Fraga, B. Schenke and R. Venugopalan for valuable discussions, and for pointing 
out several important references. This research is supported by the US DOE under contract DE-FG02-87ER40331.A008.

\bibliography{ref_cep_hbt_ms}   

\begin{thebibliography}{43}
\expandafter\ifx\csname natexlab\endcsname\relax\def\natexlab#1{#1}\fi
\expandafter\ifx\csname bibnamefont\endcsname\relax
  \def\bibnamefont#1{#1}\fi
\expandafter\ifx\csname bibfnamefont\endcsname\relax
  \def\bibfnamefont#1{#1}\fi
\expandafter\ifx\csname citenamefont\endcsname\relax
  \def\citenamefont#1{#1}\fi
\expandafter\ifx\csname url\endcsname\relax
  \def\url#1{\texttt{#1}}\fi
\expandafter\ifx\csname urlprefix\endcsname\relax\def\urlprefix{URL }\fi
\providecommand{\bibinfo}[2]{#2}
\providecommand{\eprint}[2][]{\url{#2}}

\bibitem[{\citenamefont{Itoh}(1970)}]{Itoh:1970}
\bibinfo{author}{\bibfnamefont{N.}~\bibnamefont{Itoh}}, \bibinfo{journal}{Prog.
  Theor. Phys.} \textbf{\bibinfo{volume}{44}}, \bibinfo{pages}{291}
  (\bibinfo{year}{1970}).

\bibitem[{\citenamefont{Shuryak}(1983)}]{Shuryak:1983zb}
\bibinfo{author}{\bibfnamefont{E.~V.} \bibnamefont{Shuryak}},
  \bibinfo{journal}{CERN-83-01}  (\bibinfo{year}{1983}).

\bibitem[{\citenamefont{Asakawa and Yazaki}(1989)}]{Asakawa:1989bq}
\bibinfo{author}{\bibfnamefont{M.}~\bibnamefont{Asakawa}} \bibnamefont{and}
  \bibinfo{author}{\bibfnamefont{K.}~\bibnamefont{Yazaki}},
  \bibinfo{journal}{Nucl. Phys.} \textbf{\bibinfo{volume}{A504}},
  \bibinfo{pages}{668} (\bibinfo{year}{1989}).

\bibitem[{\citenamefont{Stephanov et~al.}(1998)\citenamefont{Stephanov,
  Rajagopal, and Shuryak}}]{Stephanov:1998dy}
\bibinfo{author}{\bibfnamefont{M.~A.} \bibnamefont{Stephanov}},
  \bibinfo{author}{\bibfnamefont{K.}~\bibnamefont{Rajagopal}},
  \bibnamefont{and} \bibinfo{author}{\bibfnamefont{E.~V.}
  \bibnamefont{Shuryak}}, \bibinfo{journal}{Phys. Rev. Lett.}
  \textbf{\bibinfo{volume}{81}}, \bibinfo{pages}{4816} (\bibinfo{year}{1998}),
  \eprint{hep-ph/9806219}.

\bibitem[{\citenamefont{Aoki et~al.}(2006)\citenamefont{Aoki, Endrodi, Fodor,
  Katz, and Szabo}}]{Aoki:2006we}
\bibinfo{author}{\bibfnamefont{Y.}~\bibnamefont{Aoki}},
  \bibinfo{author}{\bibfnamefont{G.}~\bibnamefont{Endrodi}},
  \bibinfo{author}{\bibfnamefont{Z.}~\bibnamefont{Fodor}},
  \bibinfo{author}{\bibfnamefont{S.}~\bibnamefont{Katz}}, \bibnamefont{and}
  \bibinfo{author}{\bibfnamefont{K.}~\bibnamefont{Szabo}},
  \bibinfo{journal}{Nature} \textbf{\bibinfo{volume}{443}},
  \bibinfo{pages}{675} (\bibinfo{year}{2006}), \eprint{hep-lat/0611014}.

\bibitem[{\citenamefont{Bhattacharya et~al.}(2014)\citenamefont{Bhattacharya,
  Buchoff, Christ, Ding, Gupta et~al.}}]{Bhattacharya:2014ara}
\bibinfo{author}{\bibfnamefont{T.}~\bibnamefont{Bhattacharya}},
  \bibinfo{author}{\bibfnamefont{M.~I.} \bibnamefont{Buchoff}},
  \bibinfo{author}{\bibfnamefont{N.~H.} \bibnamefont{Christ}},
  \bibinfo{author}{\bibfnamefont{H.~T.} \bibnamefont{Ding}},
  \bibinfo{author}{\bibfnamefont{R.}~\bibnamefont{Gupta}},
  \bibnamefont{et~al.}, \bibinfo{journal}{Phys.Rev.Lett.}
  \textbf{\bibinfo{volume}{113}}, \bibinfo{pages}{082001}
  (\bibinfo{year}{2014}), \eprint{1402.5175}.

\bibitem[{\citenamefont{de~Forcrand}(2009)}]{deForcrand:2010ys}
\bibinfo{author}{\bibfnamefont{P.}~\bibnamefont{de~Forcrand}},
  \bibinfo{journal}{PoS} \textbf{\bibinfo{volume}{LAT2009}},
  \bibinfo{pages}{010} (\bibinfo{year}{2009}), \eprint{1005.0539}.

\bibitem[{\citenamefont{Berges and Rajagopal}(1999)}]{Berges:1998rc}
\bibinfo{author}{\bibfnamefont{J.}~\bibnamefont{Berges}} \bibnamefont{and}
  \bibinfo{author}{\bibfnamefont{K.}~\bibnamefont{Rajagopal}},
  \bibinfo{journal}{Nucl.Phys.} \textbf{\bibinfo{volume}{B538}},
  \bibinfo{pages}{215} (\bibinfo{year}{1999}), \eprint{hep-ph/9804233}.

\bibitem[{\citenamefont{Hatta and Ikeda}(2003)}]{Hatta:2002sj}
\bibinfo{author}{\bibfnamefont{Y.}~\bibnamefont{Hatta}} \bibnamefont{and}
  \bibinfo{author}{\bibfnamefont{T.}~\bibnamefont{Ikeda}},
  \bibinfo{journal}{Phys.Rev.} \textbf{\bibinfo{volume}{D67}},
  \bibinfo{pages}{014028} (\bibinfo{year}{2003}), \eprint{hep-ph/0210284}.

\bibitem[{\citenamefont{Stephanov}(2004)}]{Stephanov:2004wx}
\bibinfo{author}{\bibfnamefont{M.~A.} \bibnamefont{Stephanov}},
  \bibinfo{journal}{Prog.Theor.Phys.Suppl.} \textbf{\bibinfo{volume}{153}},
  \bibinfo{pages}{139} (\bibinfo{year}{2004}), \eprint{hep-ph/0402115}.

\bibitem[{\citenamefont{Asakawa and Nonaka}(2006)}]{Asakawa:2005hw}
\bibinfo{author}{\bibfnamefont{M.}~\bibnamefont{Asakawa}} \bibnamefont{and}
  \bibinfo{author}{\bibfnamefont{C.}~\bibnamefont{Nonaka}},
  \bibinfo{journal}{Nucl.Phys.} \textbf{\bibinfo{volume}{A774}},
  \bibinfo{pages}{753} (\bibinfo{year}{2006}), \eprint{nucl-th/0509091}.

\bibitem[{\citenamefont{Ejiri}(2008)}]{Ejiri:2008xt}
\bibinfo{author}{\bibfnamefont{S.}~\bibnamefont{Ejiri}},
  \bibinfo{journal}{Phys.Rev.} \textbf{\bibinfo{volume}{D78}},
  \bibinfo{pages}{074507} (\bibinfo{year}{2008}), \eprint{0804.3227}.

\bibitem[{\citenamefont{Fodor and Katz}(2004)}]{Fodor:2004nz}
\bibinfo{author}{\bibfnamefont{Z.}~\bibnamefont{Fodor}} \bibnamefont{and}
  \bibinfo{author}{\bibfnamefont{S.}~\bibnamefont{Katz}},
  \bibinfo{journal}{JHEP} \textbf{\bibinfo{volume}{0404}}, \bibinfo{pages}{050}
  (\bibinfo{year}{2004}), \eprint{hep-lat/0402006}.

\bibitem[{\citenamefont{Li et~al.}(2011)\citenamefont{Li, Alexandru, and
  Liu}}]{Li:2011ee}
\bibinfo{author}{\bibfnamefont{A.}~\bibnamefont{Li}},
  \bibinfo{author}{\bibfnamefont{A.}~\bibnamefont{Alexandru}},
  \bibnamefont{and} \bibinfo{author}{\bibfnamefont{K.-F.} \bibnamefont{Liu}},
  \bibinfo{journal}{Phys.Rev.} \textbf{\bibinfo{volume}{D84}},
  \bibinfo{pages}{071503} (\bibinfo{year}{2011}), \eprint{1103.3045}.

\bibitem[{\citenamefont{Nagata et~al.}(2014)\citenamefont{Nagata, Kashiwa,
  Nakamura, and Nishigaki}}]{Nagata:2014fra}
\bibinfo{author}{\bibfnamefont{K.}~\bibnamefont{Nagata}},
  \bibinfo{author}{\bibfnamefont{K.}~\bibnamefont{Kashiwa}},
  \bibinfo{author}{\bibfnamefont{A.}~\bibnamefont{Nakamura}}, \bibnamefont{and}
  \bibinfo{author}{\bibfnamefont{S.~M.} \bibnamefont{Nishigaki}}
  (\bibinfo{year}{2014}), \eprint{1410.0783}.

\bibitem[{\citenamefont{de~Forcrand et~al.}(2014)\citenamefont{de~Forcrand,
  Langelage, Philipsen, and Unger}}]{deForcrand:2014tha}
\bibinfo{author}{\bibfnamefont{P.}~\bibnamefont{de~Forcrand}},
  \bibinfo{author}{\bibfnamefont{J.}~\bibnamefont{Langelage}},
  \bibinfo{author}{\bibfnamefont{O.}~\bibnamefont{Philipsen}},
  \bibnamefont{and} \bibinfo{author}{\bibfnamefont{W.}~\bibnamefont{Unger}},
  \bibinfo{journal}{Phys.Rev.Lett.} \textbf{\bibinfo{volume}{113}},
  \bibinfo{pages}{152002} (\bibinfo{year}{2014}), \eprint{1406.4397}.

\bibitem[{che()}]{chemfo}
\bibinfo{note}{The baryon chemical potential increases with the decrease in the
  beam energy while the chemical freeze-out temperature increases with increase
  in beam energy~\cite{Cleymans:2006qe}.}

\bibitem[{\citenamefont{Lacey et~al.}(2007{\natexlab{a}})\citenamefont{Lacey,
  Ajitanand, Alexander, Chung, Holzmann et~al.}}]{Lacey:2006bc}
\bibinfo{author}{\bibfnamefont{R.~A.} \bibnamefont{Lacey}},
  \bibinfo{author}{\bibfnamefont{N.}~\bibnamefont{Ajitanand}},
  \bibinfo{author}{\bibfnamefont{J.}~\bibnamefont{Alexander}},
  \bibinfo{author}{\bibfnamefont{P.}~\bibnamefont{Chung}},
  \bibinfo{author}{\bibfnamefont{W.}~\bibnamefont{Holzmann}},
  \bibnamefont{et~al.}, \bibinfo{journal}{Phys.Rev.Lett.}
  \textbf{\bibinfo{volume}{98}}, \bibinfo{pages}{092301}
  (\bibinfo{year}{2007}{\natexlab{a}}), \eprint{nucl-ex/0609025}.

\bibitem[{\citenamefont{Csernai et~al.}(2006)\citenamefont{Csernai, Kapusta,
  and McLerran}}]{Csernai:2006zz}
\bibinfo{author}{\bibfnamefont{L.~P.} \bibnamefont{Csernai}},
  \bibinfo{author}{\bibfnamefont{J.}~\bibnamefont{Kapusta}}, \bibnamefont{and}
  \bibinfo{author}{\bibfnamefont{L.~D.} \bibnamefont{McLerran}},
  \bibinfo{journal}{Phys.Rev.Lett.} \textbf{\bibinfo{volume}{97}},
  \bibinfo{pages}{152303} (\bibinfo{year}{2006}), \eprint{nucl-th/0604032}.

\bibitem[{\citenamefont{Lacey et~al.}(2007{\natexlab{b}})}]{Lacey:2007na}
\bibinfo{author}{\bibfnamefont{R.~A.} \bibnamefont{Lacey}} \bibnamefont{et~al.}
  (\bibinfo{year}{2007}{\natexlab{b}}), \eprint{0708.3512}.

\bibitem[{\citenamefont{Hung and Shuryak}(1995)}]{Hung:1994eq}
\bibinfo{author}{\bibfnamefont{C.}~\bibnamefont{Hung}} \bibnamefont{and}
  \bibinfo{author}{\bibfnamefont{E.~V.} \bibnamefont{Shuryak}},
  \bibinfo{journal}{Phys.Rev.Lett.} \textbf{\bibinfo{volume}{75}},
  \bibinfo{pages}{4003} (\bibinfo{year}{1995}), \eprint{hep-ph/9412360}.

\bibitem[{\citenamefont{Rischke and Gyulassy}(1996)}]{Rischke:1996em}
\bibinfo{author}{\bibfnamefont{D.~H.} \bibnamefont{Rischke}} \bibnamefont{and}
  \bibinfo{author}{\bibfnamefont{M.}~\bibnamefont{Gyulassy}},
  \bibinfo{journal}{Nucl.Phys.} \textbf{\bibinfo{volume}{A608}},
  \bibinfo{pages}{479} (\bibinfo{year}{1996}), \eprint{nucl-th/9606039}.

\bibitem[{\citenamefont{Pratt}(1984)}]{Pratt:1984su}
\bibinfo{author}{\bibfnamefont{S.}~\bibnamefont{Pratt}},
  \bibinfo{journal}{Phys.Rev.Lett.} \textbf{\bibinfo{volume}{53}},
  \bibinfo{pages}{1219} (\bibinfo{year}{1984}).

\bibitem[{\citenamefont{Chapman et~al.}(1995)\citenamefont{Chapman, Scotto, and
  Heinz}}]{Chapman:1994yv}
\bibinfo{author}{\bibfnamefont{S.}~\bibnamefont{Chapman}},
  \bibinfo{author}{\bibfnamefont{P.}~\bibnamefont{Scotto}}, \bibnamefont{and}
  \bibinfo{author}{\bibfnamefont{U.~W.} \bibnamefont{Heinz}},
  \bibinfo{journal}{Phys.Rev.Lett.} \textbf{\bibinfo{volume}{74}},
  \bibinfo{pages}{4400} (\bibinfo{year}{1995}), \eprint{hep-ph/9408207}.

\bibitem[{\citenamefont{Wiedemann et~al.}(1996)\citenamefont{Wiedemann, Scotto,
  and Heinz}}]{Wiedemann:1995au}
\bibinfo{author}{\bibfnamefont{U.~A.} \bibnamefont{Wiedemann}},
  \bibinfo{author}{\bibfnamefont{P.}~\bibnamefont{Scotto}}, \bibnamefont{and}
  \bibinfo{author}{\bibfnamefont{U.~W.} \bibnamefont{Heinz}},
  \bibinfo{journal}{Phys.Rev.} \textbf{\bibinfo{volume}{C53}},
  \bibinfo{pages}{918} (\bibinfo{year}{1996}), \eprint{nucl-th/9508040}.

\bibitem[{\citenamefont{Lacey et~al.}(2014)\citenamefont{Lacey, Taranenko, Jia,
  Reynolds, Ajitanand et~al.}}]{Lacey:2013qua}
\bibinfo{author}{\bibfnamefont{R.~A.} \bibnamefont{Lacey}},
  \bibinfo{author}{\bibfnamefont{A.}~\bibnamefont{Taranenko}},
  \bibinfo{author}{\bibfnamefont{J.}~\bibnamefont{Jia}},
  \bibinfo{author}{\bibfnamefont{D.}~\bibnamefont{Reynolds}},
  \bibinfo{author}{\bibfnamefont{N.}~\bibnamefont{Ajitanand}},
  \bibnamefont{et~al.}, \bibinfo{journal}{Phys. Rev. Lett.}
  \textbf{\bibinfo{volume}{112}}, \bibinfo{pages}{082302}
  (\bibinfo{year}{2014}).

\bibitem[{\citenamefont{Lacey}(2014)}]{Lacey:2014rxa}
\bibinfo{author}{\bibfnamefont{R.~A.} \bibnamefont{Lacey}}
  (\bibinfo{year}{2014}), \eprint{1408.1343}.

\bibitem[{\citenamefont{Adare et~al.}(2014)}]{Adare:2014qvs}
\bibinfo{author}{\bibfnamefont{A.}~\bibnamefont{Adare}} \bibnamefont{et~al.}
  (\bibinfo{collaboration}{PHENIX Collaboration}) (\bibinfo{year}{2014}),
  \eprint{1410.2559}.

\bibitem[{\citenamefont{Adamczyk et~al.}(2014)}]{Adamczyk:2014mxp}
\bibinfo{author}{\bibfnamefont{L.}~\bibnamefont{Adamczyk}} \bibnamefont{et~al.}
  (\bibinfo{collaboration}{STAR Collaboration}) (\bibinfo{year}{2014}),
  \eprint{1403.4972}.

\bibitem[{\citenamefont{Aamodt et~al.}(2011)}]{Aamodt:2011mr}
\bibinfo{author}{\bibfnamefont{K.}~\bibnamefont{Aamodt}} \bibnamefont{et~al.}
  (\bibinfo{collaboration}{ALICE Collaboration}), \bibinfo{journal}{Phys.Lett.}
  \textbf{\bibinfo{volume}{B696}}, \bibinfo{pages}{328} (\bibinfo{year}{2011}),
  \eprint{1012.4035}.

\bibitem[{\citenamefont{Kisiel}(2011)}]{Kisiel:2011jt}
\bibinfo{author}{\bibfnamefont{A.}~\bibnamefont{Kisiel}}
  (\bibinfo{collaboration}{ALICE Collaboration}), \bibinfo{journal}{PoS}
  \textbf{\bibinfo{volume}{WPCF2011}}, \bibinfo{pages}{003}
  (\bibinfo{year}{2011}).

\bibitem[{\citenamefont{Ladrem and Ait-El-Djoudi}(2005)}]{Ladrem:2004dw}
\bibinfo{author}{\bibfnamefont{M.}~\bibnamefont{Ladrem}} \bibnamefont{and}
  \bibinfo{author}{\bibfnamefont{A.}~\bibnamefont{Ait-El-Djoudi}},
  \bibinfo{journal}{Eur.Phys.J.} \textbf{\bibinfo{volume}{C44}},
  \bibinfo{pages}{257} (\bibinfo{year}{2005}), \eprint{hep-ph/0412407}.

\bibitem[{\citenamefont{Palhares et~al.}(2011)\citenamefont{Palhares, Fraga,
  and Kodama}}]{Palhares:2009tf}
\bibinfo{author}{\bibfnamefont{L.}~\bibnamefont{Palhares}},
  \bibinfo{author}{\bibfnamefont{E.}~\bibnamefont{Fraga}}, \bibnamefont{and}
  \bibinfo{author}{\bibfnamefont{T.}~\bibnamefont{Kodama}},
  \bibinfo{journal}{J.Phys.} \textbf{\bibinfo{volume}{G38}},
  \bibinfo{pages}{085101} (\bibinfo{year}{2011}), \eprint{0904.4830}.

\bibitem[{\citenamefont{Miller et~al.}(2007)\citenamefont{Miller, Reygers,
  Sanders, and Steinberg}}]{glauber}
\bibinfo{author}{\bibfnamefont{M.~L.} \bibnamefont{Miller}},
  \bibinfo{author}{\bibfnamefont{K.}~\bibnamefont{Reygers}},
  \bibinfo{author}{\bibfnamefont{S.~J.} \bibnamefont{Sanders}},
  \bibnamefont{and}
  \bibinfo{author}{\bibfnamefont{P.}~\bibnamefont{Steinberg}},
  \bibinfo{journal}{Ann. Rev. Nucl. Part. Sci.} \textbf{\bibinfo{volume}{57}},
  \bibinfo{pages}{205} (\bibinfo{year}{2007}).

\bibitem[{\citenamefont{Lacey et~al.}(2011)\citenamefont{Lacey, Wei, Ajitanand,
  and Taranenko}}]{Lacey:2010hw}
\bibinfo{author}{\bibfnamefont{R.~A.} \bibnamefont{Lacey}},
  \bibinfo{author}{\bibfnamefont{R.}~\bibnamefont{Wei}},
  \bibinfo{author}{\bibfnamefont{N.}~\bibnamefont{Ajitanand}},
  \bibnamefont{and}
  \bibinfo{author}{\bibfnamefont{A.}~\bibnamefont{Taranenko}},
  \bibinfo{journal}{Phys.Rev.} \textbf{\bibinfo{volume}{C83}},
  \bibinfo{pages}{044902} (\bibinfo{year}{2011}), \eprint{1009.5230}.

\bibitem[{\citenamefont{Adare et~al.}(2013)}]{Adare:2013nff}
\bibinfo{author}{\bibfnamefont{A.}~\bibnamefont{Adare}} \bibnamefont{et~al.}
  (\bibinfo{collaboration}{PHENIX Collaboration}) (\bibinfo{year}{2013}),
  \eprint{1310.4793}.

\bibitem[{\citenamefont{Bhalerao et~al.}(2005)\citenamefont{Bhalerao, Blaizot,
  Borghini, and Ollitrault}}]{Bhalerao:2005mm}
\bibinfo{author}{\bibfnamefont{R.}~\bibnamefont{Bhalerao}},
  \bibinfo{author}{\bibfnamefont{J.-P.} \bibnamefont{Blaizot}},
  \bibinfo{author}{\bibfnamefont{N.}~\bibnamefont{Borghini}}, \bibnamefont{and}
  \bibinfo{author}{\bibfnamefont{J.-Y.} \bibnamefont{Ollitrault}},
  \bibinfo{journal}{Phys.Lett.} \textbf{\bibinfo{volume}{B627}},
  \bibinfo{pages}{49} (\bibinfo{year}{2005}), \eprint{nucl-th/0508009}.

\bibitem[{\citenamefont{Cleymans et~al.}(2006)\citenamefont{Cleymans, Oeschler,
  Redlich, and Wheaton}}]{Cleymans:2006qe}
\bibinfo{author}{\bibfnamefont{J.}~\bibnamefont{Cleymans}},
  \bibinfo{author}{\bibfnamefont{H.}~\bibnamefont{Oeschler}},
  \bibinfo{author}{\bibfnamefont{K.}~\bibnamefont{Redlich}}, \bibnamefont{and}
  \bibinfo{author}{\bibfnamefont{S.}~\bibnamefont{Wheaton}},
  \bibinfo{journal}{J.Phys.} \textbf{\bibinfo{volume}{G32}},
  \bibinfo{pages}{S165} (\bibinfo{year}{2006}), \eprint{hep-ph/0607164}.

\bibitem[{\citenamefont{Andronic et~al.}(2009)\citenamefont{Andronic,
  Braun-Munzinger, and Stachel}}]{Andronic:2009qf}
\bibinfo{author}{\bibfnamefont{A.}~\bibnamefont{Andronic}},
  \bibinfo{author}{\bibfnamefont{P.}~\bibnamefont{Braun-Munzinger}},
  \bibnamefont{and} \bibinfo{author}{\bibfnamefont{J.}~\bibnamefont{Stachel}},
  \bibinfo{journal}{Acta Phys.Polon.} \textbf{\bibinfo{volume}{B40}},
  \bibinfo{pages}{1005} (\bibinfo{year}{2009}), \eprint{0901.2909}.

\bibitem[{\citenamefont{Becattini et~al.}(2012)\citenamefont{Becattini,
  Bleicher, Kollegger, Mitrovski, Schuster et~al.}}]{Becattini:2012sq}
\bibinfo{author}{\bibfnamefont{F.}~\bibnamefont{Becattini}},
  \bibinfo{author}{\bibfnamefont{M.}~\bibnamefont{Bleicher}},
  \bibinfo{author}{\bibfnamefont{T.}~\bibnamefont{Kollegger}},
  \bibinfo{author}{\bibfnamefont{M.}~\bibnamefont{Mitrovski}},
  \bibinfo{author}{\bibfnamefont{T.}~\bibnamefont{Schuster}},
  \bibnamefont{et~al.}, \bibinfo{journal}{Phys.Rev.}
  \textbf{\bibinfo{volume}{C85}}, \bibinfo{pages}{044921}
  (\bibinfo{year}{2012}), \eprint{1201.6349}.

\bibitem[{\citenamefont{Tawfik}(2014)}]{Tawfik:2014eba}
\bibinfo{author}{\bibfnamefont{A.~N.} \bibnamefont{Tawfik}},
  \bibinfo{journal}{Int.J.Mod.Phys.} \textbf{\bibinfo{volume}{A29}},
  \bibinfo{pages}{1430021} (\bibinfo{year}{2014}), \eprint{1410.0372}.

\bibitem[{\citenamefont{Andrea~Pelissetto}(2002)}]{3DIsing1}
\bibinfo{author}{\bibfnamefont{E.~V.} \bibnamefont{Andrea~Pelissetto}},
  \bibinfo{journal}{Phys.Rept.} \textbf{\bibinfo{volume}{368}},
  \bibinfo{pages}{549} (\bibinfo{year}{2002}), \eprint{0012164}.

\bibitem[{\citenamefont{Kleinert}(1999)}]{3DIsing2}
\bibinfo{author}{\bibfnamefont{H.}~\bibnamefont{Kleinert}},
  \bibinfo{journal}{Phys.Rev.D} \textbf{\bibinfo{volume}{60}},
  \bibinfo{pages}{085001} (\bibinfo{year}{1999}).

\end{thebibliography}

\end{document}